\documentclass[twocolumn,springer,nature,superscriptaddress]{revtex4}

\usepackage{graphicx,color,hyperref}

\usepackage{amsmath}

\usepackage{marginnote}
\usepackage{dsfont}
\usepackage{amsfonts}
\usepackage{amssymb}
\usepackage{slashed}

\usepackage{braket}

\usepackage[normalem]{ulem} 

\begin{document}

\title{
Topology in the Random Scattering of Light       
}

\author{ 
Tobias~Micklitz 
}
\email{tobias.micklitz@gmail.com}
\affiliation{Centro Brasileiro de Pesquisas F\'isicas, Rua Xavier Sigaud 150, 22290-180, Rio de Janeiro, Brazil 
} 

\author{Alexander Altland}
\email{alexal@thp.uni-koeln.de}
\affiliation{Institut f\"ur Theoretische Physik, Universit\"at zu K\"oln, Z\"ulpicher Str. 77, 50937 Cologne, Germany
}

\date{\today}

\begin{abstract}

  Light scattering in random media is usually considered within the framework of
  the three-dimensional Anderson universality class, with modifications for the
  vector nature of electromagnetic waves. We propose that the linear
  dispersiveness of light introduces topological aspects into the picture. The
  dynamics of electromagnetic waves follow the same differential equations as
  those of a spin-$1$ Weyl semimetal. In the presence of disorder, this
  equivalence leads to a range of phenomena explored in this paper. These
  include topological protection against localization when helicity
  hybridization 
  is weak, the emergence of exotic phases in weakly scattering media, and
  anomalies in optical transparency in the presence of synthetic `magnetic
  fields'. We argue that some of these effects should be visible
 and investigated already in weakly disordered optical materials.

\end{abstract}

\maketitle

\vspace{.4cm}
\noindent{\bf {\large Introduction}}

The propagation of light through disordered media gives rise to complex interference 
phenomena with a wide range of practical applications, from photocatalysis~\cite{Gomard2013,Cernuto2011} 
and random lasing~\cite{Cao2005,Wiersma2008,Andreasen2011} to optical sensing~\cite{soton418245}. 
Despite conceptual parallels to the physics of scalar waves in random media,
the scattering of light remains partially enigmatic. 
One of the most intriguing theoretical predictions---three-dimensional Anderson localization 
of light under strong disorder---remains `stubbornly elusive' and experimentally unobserved. 
This persistent absence has been attributed to the vectorial nature of light~\cite{Skipetrov_2016,Yamilov2023}, 
distinguishing it from simpler scalar wave models where localization under strong 
disorder is well established~\cite{PhysRevLett.58.226,Hu2008,Flores_2013,Angel2019,PhysRevLett.101.255702,
PhysRevLett.105.090601,Billy2008,Roati2008,White2020,Schwartz2007,PhysRevLett.62.47}.

Here, we demonstrate that random scattering of light exhibits 
a complementary mechanism beyond conventional Anderson localization: topologically driven anomalies rooted in helicity conservation.
 We show that light transport in random media can exhibit robust anomalies analogous 
 to those in topological fermionic matter, 
when helicity—the projection of angular momentum along the 
direction of propagation—relaxes slowly. Our analysis relies on a formal mapping between Maxwell’s 
equations and the Schrödinger equation of a spin-1 Weyl semimetal, a class of topological systems 
recently identified in fermionic band insulators with crystalline symmetries~\cite{bradlynDiracWeylFermions2016}. 
In this analogy, 
the transverse components of light correspond to spin-$z$ projections $\pm1$, while 
longitudinal modes map to $s_z = 0$, and the Weyl cones represent left- and right-handed helicities. 
Both systems exhibit linear dispersion and a well-defined handedness, making helicity 
the classical analog of spin and a central symmetry of the problem.

 \begin{figure}[t!]
  \vspace{.6cm}
   \centering
   \includegraphics[width=0.9\columnwidth]{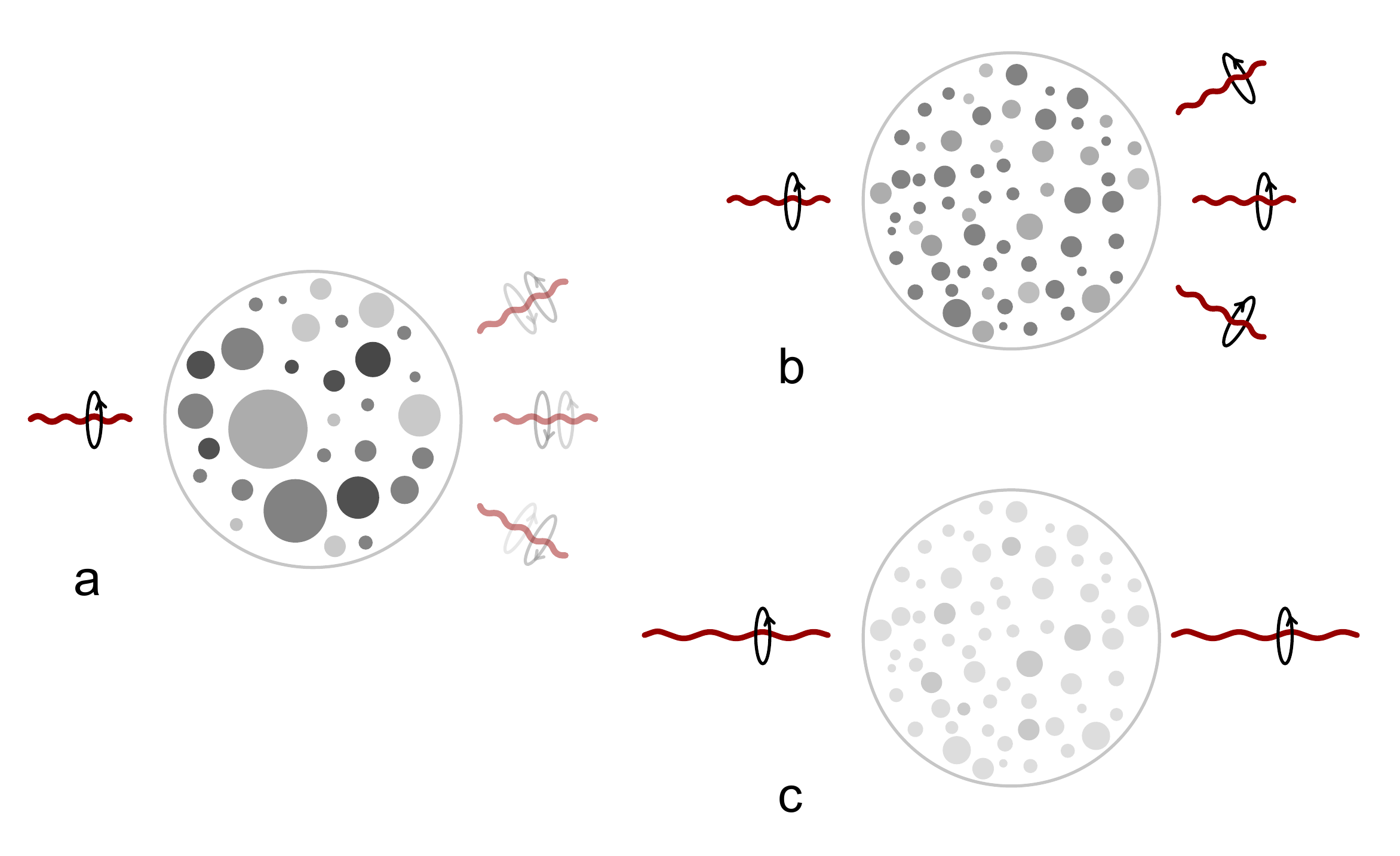}
   \caption{
   Helicity conserving media. 
   a) The scattering of light in generic optical media may alter its
   helicity, indicated by clock- or anticlockwise revolving circles. b) In
   helicity conserving media, such as synthetically engineered sub-wavelength
   dielectric spheres~\cite{Garcia-Etxarri:11,Geffrin2012}, a complete
   attenuation of transmitted signals by Anderson localization  is blocked by
   topological principles. c) Within this class of materials, there may exist a
   regime  with low but non-vanishing scattering rates where media look
   effectively transparent in the large wavenlength limit.}
   \label{fig:HelicalScattering}
 \end{figure}

In free space, the helicity of light is conserved. 
This symmetry is, however, typically broken in random optical media, where scattering 
mixes left- and right-handed circular polarizations. 
To anticipate classes of optical materials likely to exhibit 
helicity-protected transport phenomena, we recall that in random media  
the local material parameters dielectric permittivity $\epsilon_{\bold{x}}$ and magnetic permeability $\mu_{\bold{x}}$ 
define the spatially varying refractive index and impedance, 
\begin{align}
  \label{eq:RandomCoefficients}
   n_{\bold{x}}=\sqrt{ \mu_{\bold{x}} \epsilon_{\bold{x}} },
   \qquad Z_\bold{x}=\sqrt{\mu_\bold{x}/\epsilon_\bold{x}},   
\end{align} 
respectively.
Media with constant impedance, where 
$\mu_{\bold{x}} \propto \epsilon_{\bold{x}}$, preserve electric-magnetic duality, much like the vacuum, 
and thus conserve helicity even in the presence of disorder (Fig.~\ref{fig:HelicalScattering}).
Materials approaching this duality condition, such as 
sub-wavelength dielectric spheres~\cite{Garcia-Etxarri:11,Geffrin2012} and high-index semiconductors, including silicon, 
germanium, and rutile-TiO$_2$~\cite{Geffrin2012}, exhibit near-conservation of helicity over specific 
frequency ranges, as demonstrated by persistent circular polarization in scattering 
experiments. These systems realize what is known as the `first Kerker condition'~\cite{Kerker:83} and 
provide fertile ground for observing topology-driven transport phenomena in optics.

In generic disordered media, where helicity is not strictly conserved, these anomalies become fragile. 
Nevertheless, as long as helicity mixing remains weak, light propagation retains an anomalous 
character over intermediate length scales. A spectral flow mechanism~\cite{PhysRevB.93.075113,PhysRevX.14.011057}, 
analogous to that in Weyl semimetals, continues to be at work, 
and raises key questions: how far do the anomalies extend, and what signatures do they leave in measurable optical observables?
To address these questions, we develop an effective field theory of light transmission in disordered media, 
that captures the physics beyond the scattering mean free path. 
The theory generalizes the Chern-Simons (CS)
framework developed for spin-1/2 Weyl fermions~\cite{Altland2015,PhysRevB.93.075113} to spin-1 
electromagnetic fields. It identifies two central parameters: a bare transmission coefficient $\sigma$, 
derived within a self-consistent Born approximation, and a golden rule 
helicity-mixing rate $\Gamma_{\textrm{h}}$, 
which quantifies coupling between opposite helicities. 
Together, these parameters define a nonlinear $\sigma$-model (cf.
Eqs.~\eqref{eq:DiffusionAction} and Eqs.~\eqref{eq:CouplingAction} below), 
generalizing the field theory of light pioneered by
John~\cite{PhysRevLett.58.2486,PhysRevLett.53.2169,PhysRevB.31.304} to
individually resolved helical sectors.
 Crucially, each sector contributes a topological Chern-Simons term to the action 
 (Eq.~\eqref{eq:chern_simons}), 
which enforces non-vanishing transmission and blocks localization. 
This anomaly persists up to a crossover 
 length scale $l_\textrm{h}\sim
\sqrt{\sigma/\Gamma_\textrm{h}}$, 
beyond which helicity mixing restores conventional, diffusive transport.

Building on Chern-Simons theory, we uncover three qualitatively distinct transport regimes, 
each governed by the level of disorder and the extent of helicity conservation. 
These include a topological suppression of Anderson localization, anomalous 
transport analogous to effects in Weyl semimetals, and a transition to an effectively 
transparent phase at weak disorder. Together, they reveal a class of phenomena 
inaccessible to scalar wave theories and underscore the need for a topology-aware 
framework—an approach we develop and apply in the following sections.

\begin{figure}[t!]
  \centering
  \includegraphics[width=1\columnwidth]{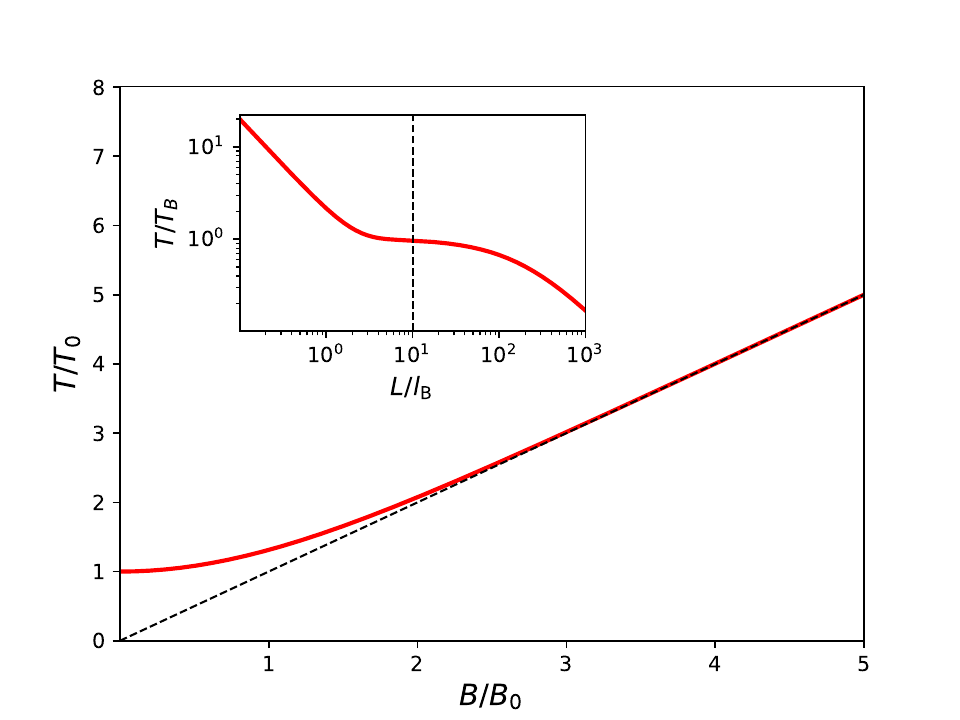}
  \caption{
  Transmission, $T$, vs. magnetic field $B$. Here  $T_0=R^2\sigma/(8L)$ and $B_0=\sigma/(4L)$, 
  and assuming $l_{\rm h}>L$. For $B>B_0$ the field dependence becomes asymptotically linear.
  Inset: $T$ vs. length $L$ in units of $l_{\rm B}$, with $l_{\rm h}=10l_{\rm B}$ 
  (indicated by the vertical dashed line) 
  and $T_B=R^2 B/2$. For further details on the parameters, see discussion 
   below Eq.~\eqref{eq:CouplingAction}.
   }
  \label{fig:transmission}
\end{figure}

\vspace{.4cm}
\noindent{\bf {\large Results and Discussion}}

We now examine the three disorder-dependent transport regimes in detail. 
Each reflects a unique topological 
response captured by the Chern-Simons theory framework—and each is 
missed by conventional scalar descriptions of light scattering.

\textbf{Strong `dual' disorder: Absence of Anderson localization.} Light 
scattering in the limiting case of helicity-conserving media is governed by 
 principles otherwise protecting extended surface states of topological
insulators or Weyl semimetals. This correspondence manifests itself in 
the Chern-Simons  theory  outlined above,  
and implies the absence of  Anderson localization in sectors of definite helicity. 
We note that in the optical context, this protection mechanism engages longitudinal electromagnetic modes, which are absent 
in both vacuum and electronic systems. This finding underscores the importance of longitudinal modes for 
light delocalization,  highlighted from a different perspective in Ref.~\cite{PhysRevB.103.174204} 
 and demonstrated in the recent numerical simulations~\cite{Yamilov2023}.

\textbf{Intermediate disorder: Anomalous transport.} Weyl semimetals show various anomalies in
 electronic transport, among them the  \emph{chiral magnetic effect}---an electric current driven by application of a static magnetic field. 
  The analog of a magnetic field required for its optical variant, the \emph{helical vortical effect}
 ~\cite{Hayata2018,PhysRevD.78.074033,PhysRevLett.103.191601}, has been proposed
 for  meta-materials, such as resonator lattices~\cite{Lin2014,Fang2012} 
  and magnetoelectric materials~\cite{Hayata2018}. 
 We here consider their realization in systems in rotational motion~\cite{PhysRevA.96.043830}, such as cylinders of
 radius $R$ and length $L$, rotated around their axis at uniform angular velocity
 $\Omega$.
 In the rotating frame, Maxwell's equations for waves of frequency $\omega$
  become equivalent to the Schrödinger equation for a spin-1 Weyl fermion subject
  to a magnetic field $B=\frac{1-\epsilon\mu}{c^2}\omega \Omega$ along the
  cylinder axis. In the presence of disorder, the Chern-Simons theory applied to this geometry predicts a
  succession of three qualitatively different regimes, marked by length scales
  $l_B<l_\textrm{h}$: For cylinder lengths $L<l_B\equiv \sigma/(8B)$, the
  transmission is Ohmic with a length dependence $T\simeq \pi R^2\sigma/L$. 
  Here, $\sigma$ and $T$ are the optical analog of the medium's
  conductivity and conductance, respectively. At the scale $l_B$ a crossover
  into a regime with length independent transmission $T=N_\phi/2\pi$ occurs,
  where $N_\phi=\pi R^2 B$ is proportional to the magnetic field and plays a role
  analogous to the number of ``flux quanta'' threading the system. Physically,
  this crossover reflects a change from diffusive light propagation at short
  scales to the \emph{drift} of two transport channels of opposite helicity at
  larger scales. Remarkably, this mechanism leads to 
  immunity  against Anderson localization, including in elongated quasi-one 
  dimensional geometries where localization is generally strong and occurs 
  even for weak disorder. Drift transport prevails until helicity mixing at
  the scale $l_\textrm{h}$ leads into another regime of Ohmic $T\sim 1/L$, and
  eventually localized $T\sim \exp(-L/\xi)$ transport.  
 These  regimes may be observable in scattering experiments monitoring light
 transmission either as a function of length or, perhaps more conveniently,
 sample rotation frequency, see also Fig.~\ref{fig:transmission}.

 \textbf{Weak disorder: Effectively clean phases.} The construction of the field
 theory outlined above  relies on a finite self consistently obtained scattering
 rate $\kappa$. Pioneering work~\cite{PhysRevB.33.3263,PhysRevB.100.125160} has
 shown that in the long wavelength limit, $\lambda\to \infty$, there exists a 
 quantum critical point marked by a disorder strength $w=2\pi^2/\Lambda$ below which 
 the self consistent scattering rate \emph{vanishes}. Here, $\Lambda^{-1}$ is a small 
 distance cutoff, to be identified with
 the correlation range of the disorder (see below). For lower concentrations,
 renormalized perturbation theory~\cite{PhysRevB.33.3263} around the Weyl nodal
 point reveals that scattering  becomes irrelevant: long wavelength
 propagation modes effectively average over medium fluctuations and pass it
 unhindered. 
 For finite wavelengths, a fan-like
 critical region emerges above the critical point. 
 Helicity conserving optical (meta)materials may provide the required
 homogeneity levels to realize the weak scattering universality class, where a
 system looks ``effectively transparent'' in the long wavelength limit. They may
 also define an ideal platform to study its behavior under changes of disorder
 concentration or light frequency. 
 We finally
 note that the existence of a quantum critical point between two phases
 at low disorder and large wavelength is a specific feature of the Weyl
 universality class, different from the generic feature that scattering becomes
 less effective in long wavelength limits~\cite{hulst1981light,10.1063/1.881300}.


\textbf{Microscopic analysis.} In the following, we outline the derivation of the results
above,  starting from the source-free Maxwell equations for electromagnetic waves 
with frequency $\omega$ (summation convention) 
\begin{align}
\label{eq:Maxwell1} 
\partial_i D_i
&=0,
\qquad
\epsilon_{ijk}\partial_j E_k
=\frac{i\omega}{c} B_i,
\\
\label{eq:Maxwell2}
\partial_i B_i
&=0,
\qquad
\epsilon_{ijk}\partial_j H_k
=-\frac{i\omega}{c} D_i.
\end{align}
The presence of a scattering medium enters through the 
 relations $B_i=\mu H_i$, $D_i=\epsilon E_i$ where the magnetic
 permeability $\mu_{\bold{x}}$  
 and permittivity 
 $\epsilon_\bold{x}$ are randomly fluctuating, and $c$ is 
 the velocity of light in vacuum.

 The standard approach to wave
 localization~\cite{PhysRevLett.53.2169,LAGENDIJK1996143} combines
 Eqs.~\eqref{eq:Maxwell1} and \eqref{eq:Maxwell2} to a single random wave
 equation for the electric field, resembling a Schrödinger equation in a random
 potential. However, following Ref.~\cite{BIALYNICKIBIRULA1996245}, we here work in an
 alternative representation, emphasizing two crucial aspects of light: linear
 dispersiveness and helicity. To this end, we introduce the rescaled field
 vectors  ${\cal E}_i= \sqrt{\epsilon} E_i$ and
 ${\cal H}_i= \sqrt{\mu} H_i$, along with the left and right circular polarized
 amplitudes $\Phi^\pm=({\cal E}\pm i {\cal H})/\sqrt{2n}$, where
 $n=\sqrt{\epsilon \mu}$ the (local) refractive index. It is then
 straightforward to show (see methods and Ref.~\cite{BIALYNICKIBIRULA1996245}) 
 that the six-component field vector
 $\Phi=(\Phi^+,\Phi^-)$  obeys
 the Dirac-like equation 
\begin{align}
\label{eq:Dirac}
\left( 
-i\tau_3\otimes c\slashed{\partial}
-
\tau_1\otimes c\slashed{a}_{Z}
+
V
\right)
 \Phi
&=
\omega\Phi.
\end{align}
Here,  the matrices $\tau_i$ act between the helical subspaces of the
\(\Phi\)-vector and  we use a variant of the Feynman slash notation,
$\slashed{v}=v_iS_i$, where $(S_j)_{\alpha\beta}= -i\epsilon_{j\alpha \beta}$
are  three-dimensional anti-symmetric matrices acting on the helical components
$\Phi^\pm$ of the field vector separately. Their commutation relations $[S_i,
S_j]=i\epsilon_{ijk} S_k$ and the normalization $S_iS_i=2\openone_3$ identify
them as spin-1 operators  heralding the nature of the photon that would
emerge upon quantization of the theory~\cite{BIALYNICKIBIRULA1996245,PhysRevD.96.051902}. 
Randomness enters the problem through the scalar potential
$V_\bold{x}=\omega(1-n_\bold{x})$ and vector potential $\slashed{a}_{Z,\bold{x}}= \frac{1}{2}
\slashed{\partial} \ln Z_\bold{x}$.

In the absence of
impedance fluctuations, $Z_\bold{x}=\textrm{const.}$, or $\slashed{a}=0$, the
equation splits into two blocks, 
\begin{align}
  \label{eq:WeyEquation}
\left(\mp i c\slashed{\partial}+ V\right) \Phi^\pm= \omega \Phi^\pm,
\end{align} 
describing the independent propagation of fields of opposite helicity. Upon
rescaling by $\hbar$, it becomes identical to the  Weyl  equation of spin-$1$
states~\cite{bradlynDiracWeylFermions2016,Lv2017,PhysRevA.96.033634}. In
condensed matter physics, spin-$1$ Weyl states have been identified as low
energy quasi-particles featuring at the band crossing points of crystals with
nontrivial space group symmetries~\cite{bradlynDiracWeylFermions2016}. In the
clean case, $V=0$, Eq.~\eqref{eq:WeyEquation}  describes two linearly dispersive
bands, plus  a flat band of zero energy quantum states. The two nontrivial bands
carry an integer topological index $C=\pm 2$  obtained, e.g., by integration of
the Berry curvature of eigenstates $\Phi^\pm$ in momentum-space over closed
surfaces surrounding $p=0$. The presence of these quantum numbers is at the
heart of topological protection against localization in the presence of
disorder. In optics, the topologically charged bands represent transverse modes
of the electromagnetic field, labeled by their propagation wave-vector, and the
degenerate band corresponds to longitudinal modes. While the latter are
irrelevant in vacuum, they open a transmission channel in random media that
becomes effective at sub-wavelength scales, thereby hindering
localization~\cite{PhysRevB.103.174204,Yamilov2023}. As we will discuss next,
this mechanism draws upon the
interplay between transverse and longitudinal fluctuations, highlighting their
interplay in random light scattering.

\textbf{Chern-Simons theory.} In quantum physics, the multiple wave 
interference processes culminating in 
Anderson localization at large length scales are  described by field theories
whose degrees of freedom, $T(\bold{x})$, are matrices reflecting the symmetries
of a given problem~\cite{Efetov-book,MIRLIN2000259}. The transfer of these
methods to light scattering was pioneered
in~\cite{PhysRevLett.58.2486,PhysRevLett.53.2169,PhysRevB.31.304}, with the
prediction that  length scales exceeding the mean free scattering path, $\kappa^{-1}$,  are described
by the action
\begin{align}
  \label{eq:DiffusionAction}
  S_0[Q]= \frac{\sigma}{16}
\int d^3 x\, \textrm{tr}(
\partial Q\partial Q),
\end{align}
where $Q=T \sigma_3 T^{-1}$, here and throughout unlabeled Pauli matrices 
$\sigma_i$ act in a two-dimensional multiplet space  distinguishing between
advanced and retarded wave propagation, and the  constraint
$Q^2=\mathds{1}$ identifies the action as that of a nonlinear $\sigma$-model.
The coupling constant  $\sigma =
(\omega^2+3\kappa^2)\left(\omega^2-\frac{1}{3}\kappa^2\right)/\left(
\pi\kappa(\omega^2+\kappa^2)\right)$  depends on frequency and $\kappa^{-1}$ (see the methods section), 
and it determines  the optical transmission
properties at large length scales: depending on whether $\sigma\kappa^{-1}$ is
large or small compared to unity, Eq.~\eqref{eq:DiffusionAction} describes the flow into a
diffusive or a localized phase, the two scenarios being separated by the
Anderson localization transition~\cite{RevModPhys.80.1355}.

Our resolution of distinct helical sectors changes this result. Instead of a
single slow field, we now have two, $Q_\pm$, describing propagation in channels
of opposite helicity via actions $S_\pm[Q_\pm]=S_0[Q_\pm]\pm S_\textrm{top}[A_\pm]$.
Alongside the standard gradient action $S_0[Q_\pm]$, these contain a topological
action of Chern-Simons form
$S_\textrm{top}[A_\chi]=S_\textrm{CS}[A^+_\chi]-S_\textrm{CS}[A^-_\chi]$, where
\begin{align}
  \label{eq:chern_simons}
  S_{\rm CS}[A]
  &=
  -\frac{i\epsilon_{ijk}}{8\pi}
  \int d^3x\,
  {\rm tr}\left(
  A_i  \partial_j A_k  
  +
  \frac{2}{3} A_i A_j A_k 
  \right), 
  \end{align}
  $A_\chi^s=
T^{-1}_\chi \partial T_\chi P^s$, and $P^s=
\frac{1}{2}(\mathds{1}+s\sigma_3) $ are projectors on 
retarded or advanced propagation channels. 

To rationalize the emergence of this  term, consider the unitaries $T_\pm$ as
nonabelian gauge transformations. Our theory then resembles that  of
three-dimensional Dirac fermions  coupled to  vector potentials
$\bold{A}=T^{-1}\partial T$. This system is subject to the parity
anomaly~\cite{PhysRevD.29.2366}, where  ultraviolet singularities require giving
up parity symmetry in exchange for maintained gauge invariance. On general
grounds,  this principle requires the presence of a Chern-Simons term in gauge
field actions.

\begin{figure}[t!]
  \centering
   \includegraphics[width=0.5\columnwidth]{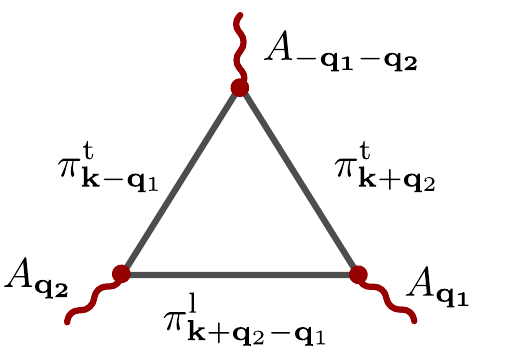}
  \caption{ Triangle scattering amplitude. 
  A Feynman diagram contributing to the $A^3$-term in
  the Chern-Simons action Eq.~\eqref{eq:chern_simons}. 
  Three slow fields $A$ of low
  characteristic wave-vectors $\bold{q}_i\ll \Lambda$ are connected by electromagnetic
  field propagators, one of which needs to be a longitudinal mode. The integral over wave-vectors $\textbf{k}$ is
  dominated by large $|\textbf{k}|\sim \Lambda$, where the influence of disorder on
 the latter is exceptionally strong. }
  \label{fig:FeynmanDiagram}
\end{figure}

Referring for to the methods section for details, the CS term reflects the presence of 
``triangle'' scattering amplitudes~\cite{PhysRevD.29.2366}, coupling three photon propagators to
the gauge field, see Fig.~\ref{fig:FeynmanDiagram}. We find that these amplitudes necessarily involve the coupling of transverse
photon propagators to spin $S_z=0$ longitudinal propagators. What is more, the
dominant contributions  come from high-lying wave-vectors
of the order of the UV-cutoff $\Lambda$, 
which is the parameter region dominantly supporting longitudinal propagators. 
These observations underscore the importance of longitudinal modes in the 
physics of light delocalization,  pointed out from a different perspective in~\cite{PhysRevB.103.174204,Yamilov2023}.  

The derivation of this effective theory answers the questions raised above. The
low energy theory, split into two helical sectors
$S[Q_+,Q_-]=S_+[Q_+]+S_-[Q_-]$, is mathematically identical to that of Weyl
quantum matter in the absence of inter-node
scattering~\cite{Altland2015,PhysRevB.93.075113}. An indpepedent way of rationalizing 
its protection against localization makes reference to the bulk-boundary principle 
of topological quantum matter: individual of the Chern-Simons  actions $S_\pm$ describe isolated surfaces  of
four-dimensional topological insulators~\cite{Ryu_2010}, where localization is
excluded via a spectral flow principle~\cite{PhysRevX.14.011057}. Our derivation
of the CS action generalizes this result, previously established for spin-$1/2$
states, to spin-$1$. Interestingly, the protection mechanism extends to quasi-one dimensional 'waveguide'
geometries~\cite{PhysRevB.93.075113}, where localization would otherwise be
unavoidable regardless of the value of $\sigma$. 

To find out what happens  once impedance
$Z$-fluctuations couple the helicity sectors, we need to be  more
explicit concerning the realization of randomness.   
Assuming  scattering centers smaller than the wavelength,
 $\Lambda^{-1}<c/\omega$, we consider Gaussian variances as in the first two
 rows of table~\ref{table:1}, with $\delta$-functions defined to be smeared over
 scales $\Lambda^{-1}$. The light scattering rate $\kappa(w,\Lambda,\omega)$,
 and the light transmission $\sigma=\sigma(\omega,\kappa)$ depend on these
 parameters  in a manner that can be computed via self-consistent Born
 approximation, but will not be essential for us. Presently, the most important
 parameter is a \emph{helicity coupling constant} $\Gamma_\textrm{h}$ whose
 dependence on impedance fluctuations and the
 density of scattering modes is stated in Table~\ref{table:1}.
 This constant enters the  theory via the  term  
\begin{align}
  \label{eq:CouplingAction}
  S_\textrm{c}[Q]=\Gamma_\textrm{h}\int d^3x\,\textrm{tr}(Q_+ Q_-),
\end{align}
coupling the two helical  fields. At large length scales, Eq.~\eqref{eq:CouplingAction}
enforces a locking $Q_+=Q_-\equiv Q$. Inspection of the sign structure of the
CS terms shows that they cancel out in this limit. 

To parametrically estimate the  corresponding crossover scale $l_\textrm{h}$, we
compare the characteristic value of the diffusion action $\sim \sigma l^3/l^2$
to that of the helicity action $\sim \Gamma_\textrm{h} l^3$ at length scales
$l$. Equating these values, we obtain $l\equiv l_\textrm{h}\sim
\sqrt{\sigma/\Gamma_\textrm{h}}$. Above this length scale, the theory will flow
into an Anderson localized phase, provided the bare transmission coefficient
$\sigma(\omega,\kappa)$ is supercritically weak. 

Generalized for the presence of a synthetic magnetic field $B$,  the theory
defined by Eqs.~\eqref{eq:DiffusionAction},\eqref{eq:chern_simons}, and
\eqref{eq:CouplingAction}  predicts a number of intriguing  transport phenomena.
Referring to Ref.~\cite{PhysRevB.93.075113} for details, the vector potentials
entering the CS term $A_i\to A_i + A_{\textrm{e},i}$ must then be
interpreted as the sum of the fluctuation fields and the  potential
$A_{\textrm{e},i}$ representing $B$. In this case, it becomes a one-derivative
operator describing \emph{drift} along
quasi one-dimensional geometries aligned with the field~\cite{PhysRevB.93.075113}. A 
dimensional estimate shows that for lengths  exceeding the crossover scale $l_B$
 the drift action dominates over the diffusive
Eq.~\eqref{eq:DiffusionAction} causing a  crossover from `Ohmic' to constant
transmission (see Fig.~\ref{fig:transmission}).  Only at scales
$L>l_\textrm{h}$, the helical drift terms cancel out, and we re-enter an Ohmic,
and eventually an Anderson localized regime. We here assumed a hierarchy of
scales $l_B<l_\textrm{h}$ corresponding to weak helicity breaking. In the
opposite case, the CVE will be unobservable in a disordered medium.

\begin{table}[t!]
\begin{tabular}{l|l}
\hline

{\rm refractive index }& $\langle
V_\bold{x}V_{\bold{x}'}\rangle= wc^2 \delta(\bold{x}-\bold{x}')$  \\
{\rm impedance } &
$\langle {a}_{i,\bold{x}}a_{j,\bold{y}} \rangle= \gamma \delta(\bold{x}-\bold{y})\delta_{ij}$
\\
\hline
scattering rate & $\kappa=\kappa(w,\Lambda,\omega)$\cr
transmission coefficient & $\sigma=\sigma(\omega,\kappa)$\cr 
helicity coupling\,\,\, & $\Gamma_\textrm{h}=6 \gamma \left(\kappa/wc\right)^2$\cr 
helical length & $l_\textrm{h}\sim (\sigma/\gamma)^{1/2}wc/\kappa$\cr
\hline
  \end{tabular}
\caption{ Material parameters. Parameters derived from variances of the potential, $V$
and vector potential, $a_i$, generated by the refractive index, respectively, gradient of the impedance. 
See also Eqs.~\eqref{eq:RandomCoefficients} and around Eq.~\eqref{eq:Dirac}  for definitions and further   
 discussion.
 }
\label{table:1}
\end{table}

{\bf Experimental realizability.} Before concluding, let us speculate  
on the experimental accessibility of the phenomena discussed above. Again, the
situation calls for a separate consideration of strong, intermediate and weak
disorder. Beginning with the latter, weakly disordered meta-materials with approximately
dual scatterers have been theoretically~\cite{PhysRevLett.114.113902} and
experimentally studied~\cite{NegoroDualScatterers}. 
Specifically, the 
`first Kerker condition'~\cite{Kerker:83} 
is realized in dielectric
sub-wavelength spheres whose scattering profile was
found~\cite{Garcia-Etxarri:11,Geffrin2012} to be indistinguishable from that of
dual magneto-dielectric spheres. Silicon (and other semiconductor materials such
as germanium and rutile-TiO$_2$~\cite{Geffrin2012}) are likewise  close to  a
near-dual limit for certain frequency ranges of the scattering light.
Refs.~\cite{PhysRevB.40.9342,PhysRevE.49.1767,Gorodnichev_2017} noted the
persistence of circular polarization indicative of a dual limit in resonant
dielectric Mie particles with high magnetic polarizability.

The limit of strong helicity conserving scattering, on the other hand, may be
elusive in these present day materials. For one, the Kerker condition holds only for narrow
frequency ranges, defined by specific
ratios between scatterer size and light wavelength, thus constraining  the size
distribution of scatterers in random media, cf.~\cite{PhysRevLett.114.113902}. Second, higher-order
multipolar responses compromising helicity may become effective via proximity effects
in closely packed materials. In view of this situation, the weak and
intermediate regime  appear to be most accessible at present. 

We suggested probing this regime in  cylindrical geometries rotating with
angular velocity $\Omega$. Assuming
sample radii $R$ in the centimeter range, and transmission coefficients $T=\pi
R^2\frac{\sigma}{L}$ of the order of unity, the crossover length $l_B\sim
\frac{\sigma c^2}{8\omega\Omega}$ will be comparable to the system size for rotation
frequencies down by  a factor $(\lambda/R)^2$ relative to the frequency
of light, $\omega$. For example, for  wavelength in the visible range, $\lambda\sim 500
\textrm{nm}$, and including numerical factors, we obtain characteristic
frequencies $\Omega$ in the $\textrm{kHz}$ range, which may be technically
achievable. We also note that in the rotating frame an incoming cylindrically
symmetric wave front will be perceived as stationary, showing that effects
related to the scattering off non-stationary scatterers can be avoided in this
setting. However, in view of the remaining experimental challenges, large
scale  numerical simulations as in Refs.~\cite{Hayata2018,Lin2014,Fang2012,PhysRevD.78.074033,PhysRevLett.103.191601} 
will be natural first steps in targeting topological phenonmena in the weak and
intermediate disorder regime.

\vspace{.4cm}
\noindent{\bf {\large Conclusions}}

The scattering of light in random media is commonly discussed
within the framework of the Anderson universality class, with additional account
for the vectorial nature of electromagnetic waves. Here, we
have shown that the linear dispersion of electromagnetic waves, combined with
its vector nature, implies more far-reaching departures from standard
localization theory. It puts the  problem into the universality class of the
random spin-$1$ Weyl semimetal, where  the role of Weyl cones of opposite
chirality is taken by waves of opposite helicity. As a  consequence the
transmission of light will show sensitivity to the degree of helicity
mixing. We introduced Chern-Simons theory as a universal framework
for the description of the ensuing phenomena at length scales exceeding the
scattering mean free path. We used this platform to discuss three
manifestations of the equivalence to the quantum Weyl problem: Topological
protection against localization in the limiting case of strong helicity
preserving disorder, transmission anomalies caused by an interplay of
intermediate strength
disorder and the chiral vortical effect, and the emergence of ``effectively clean'' phases in which the presence
of a weak concentration of scattering centers becomes irrelevant in a quantum
phase transition. In view of the comparatively
high level of control available in optical meta-materials, the observability of
such phenomena may be less challenging than in the  exotic crystalline quantum
materials  harboring condensed matter realizations of spin-$1$ Weyl cones. Perhaps, 
light scattering may even become an efficient macroscopic analog
simulator of universal  phenomena predicted, but so far unobserved in Weyl
quantum transport.

\vspace{.4cm}
{\noindent\bf {\large Methods}}

Throughout this section, we set $c=1$ for notational simplicity. 

{\bf `Dirac' equation for photons.} We start by defining 
the six component vector $\Psi=E\oplus H$, and the diagonal matrix $D=\epsilon \mathds{1}\oplus \mu \mathds{1}$, to write Maxwell's equations as 
\begin{align*}
  i\partial_t D \Psi = \tau_2 i \slashed{\partial}\Psi, 
\end{align*}  
where $\tau_2$ acts in the electric/magnetic subspace of $\Psi$, and the action of  the differential operator on three-component vectors is defined through the slash notation $(\slashed{\partial}X)_k = i \epsilon_{klm}\partial_l X_m$, or $\slashed{\partial}X = S_l \partial_l X$, with the hermitian matrices $(S_l)_{mk}\equiv i \epsilon_{mlk}$. Assuming harmonic time dependence of $\Psi$ with characteristic frequency $ \omega$, the equation reduces to  $\omega\Psi = D^{-1} \tau_2 (-i \slashed{\partial})\Psi$. 
We next apply a similarity transformation, $\Psi = D^{-1 /2}\tilde{\Phi}$, to obtain the representation 
\begin{align*}
  \omega \tilde{\Phi}=D^{-1 /2} \tau_2 (-i \slashed{\partial}) D^{-1 /2}\tilde{\Phi}.
\end{align*} 
The operator on the r.h.s. is manifestly hermitian (safeguarding the existence of solutions with real frequencies $\omega$) but inconvenient to work with as the scattering disorder is hiding in the factor matrices $D$. We aim to bring it into a more customary form `derivative operator + disorder potential'. 
To this end, we first  represent the matrix $D$ in terms of the refractive index and impedance 
as $D=n \exp(\ln Z \tau_3)$, leading to 
\begin{align*}
  \omega n \Phi 
  &=  \left(-i \slashed{\partial}\tau_2 - \frac{\tau_1}{2}(\slashed{\partial}\ln Z)\right)\Phi, 
\end{align*}
where $\Phi = n^{-1 /2}\tilde \Phi$. Finally, with $V=\omega(1-n)$, and
$\slashed{a}_Z\equiv \frac{1}{2}\slashed{\partial}\ln Z$  we obtain the desired form of the equation, 
\begin{align*}
  \omega \Phi =  \left(-i \slashed{\partial}\tau_2 - \tau_1\slashed{a}_{Z,\textbf{x}}+V_\textbf{x}\right)\Phi. 
\end{align*}
Performing a unitary rotation around $\tau_1$ 
to the helical eigenstates $\Phi \to (\Phi^+,\Phi^-)$, we arrive at Dirac-like equation stated in the text.

{\bf Replica field theory.} We here provide details concerning the derivation of
the Chern-Simons field theory discussed in the main text. Referring to
Ref.~\cite{PhysRevB.93.075113} for a more detailed exposition our discussion
will be succinct but self-contained. Specifically, we highlight departures from
the theory of the spin $1/2$-system. Alongside the presence of longitudinal
modes, these include a different behavior under physical time reversal: while
for spin-$1/2$  time reversal squares to minus the identity, it is even in
the spin-$1$ case. This puts our problem into symmetry class AI
(otherwise known as the Wigner-Dyson orthogonal class) in the nomenclature of
Refs.~\cite{altlandNonstandardSymmetryClasses1997,Ryu:2010fk}. 
A detailed account of the calculations including all intermediate steps can be found 
in the Supplementary Notes 1 and 2.

The starting point of the field theory description 
is the replica partition function, 
\begin{align}
\label{app_eq:partition_function}
{\cal Z}[j]
&=
\int D\psi 
\left\langle
e^{i \int d^3x\, \bar{\psi}
(i\delta \sigma_3 + \omega - \tau_3\otimes\slashed{k}-V)  \psi
}
\right\rangle_{\rm dis}, 
\end{align}
where $\psi=\{ \psi_{s,\sigma,a,\nu,\chi}(\bold{x})\}$ is a 
$(2\times 2\times R \times 3 \times 2)$ component Grassmann field. Here, $s=1,2$ is 
a two-component index distinguishing between retarded and advanced fields,  
$a=1,\ldots,R$ is a replica index, $\nu=1,2,3$ labels the spin degree of freedom, 
 and $\chi=\pm$  helicity. 
The last remaining index, $\sigma=1,2$,  accounts for time-reversal symmetry, which is realized via the relation  
$\bar{\psi}=(i\sigma^{\rm tr}_2\psi)^T$. Finally, $\sigma^{\rm tr}_i$ and
$\sigma_i$ are Pauli matrices in the spaces of $\sigma$- and
$s$-indices, respectively. 

The action in Eq.~\eqref{app_eq:partition_function} possesses a continuous
symmetry which will be crucial for all what follows: it is invariant under
uniform rotations $\psi\rightarrow T\psi$, $\bar \psi\to \bar \psi T^{-1}$
commuting with  $\tau_3\otimes\slashed{k}$. These are spatially uniform
rotations structureless in spin-indices $\nu=1,2,3$, and diagonal in helical
$\tau$-space. Consistency with the time reversal structure relating
$\psi$ and $\bar \psi$ implies the further constraint $T^T=\sigma_2^{\rm
tr}T^{-1}\sigma_2^{\rm tr}$, identifying $T$ as an element of the symplectic group
$\textrm{Sp}(4R)$.

{\it Hubbard-Stratonovich decoupling:---}The Gaussian average over 
fluctuations $V$ in Eq.~\eqref{app_eq:partition_function} generates  a quartic `interaction term', local in space but with tensor structure  in the `internal' indices $s,\sigma,a,\nu,\chi$.
We decouple this term by a matrix-field $Q=\{Q_{\chi\chi',\nu\nu'}^{ss',\sigma\sigma',aa'}\}$ via Hubbard-Stratonovich transformation and subsequently perform the quadratic integral over $\psi$ to obtain 
the matrix field partition sum  ${\cal Z}=\int dQ e^{-S[Q]}$, with
\begin{align*}
S[Q]
&=
\frac{1}{2w}{\rm Tr}\, Q^2
-
\frac{1}{2}{\rm Tr}\ln \left(\omega - \tau_3\otimes \slashed{k} + i\delta\sigma_3  + i Q
\right). 
\end{align*}

\emph{Mean-field analysis:---}Variation of the action in $Q$ 
leads to the saddle point equation, 
\begin{align}
Q&=
\frac{iw}{2} \int^\Lambda (dk)\, 
\frac{1}{\omega - \tau_3\otimes \slashed{k} + i\delta\sigma_3 + i Q},
\end{align}
where $(dk)\equiv d^3k/(2\pi)^3$ and the superscript $\Lambda$ indicates that the, 
formally divergent, momentum integration is to be cut-off at the inverse of 
the correlation length $\Lambda^{-1}$ implicit in the definition of our 
disorder correlators. We assume this to be the smallest length scale in the problem, 
in particular $\Lambda^{-1}\ll c /\omega$. 

The structure of the equation indicates that
the mean field $Q$ plays the role of a self-consistent Born ``self energy'' of light
propagators picked up upon scattering. The detailed solution of this equation
will lead to a self energy operator with structure in spin space ($\nu$),
reflecting differences between longitudinal and transverse self energies. While
these are important for the quantitative solution computation of scattering
cross-sections, they have no qualitative significance for our analysis. We
therefore seek an approximate solution in terms of the spin-isotropic ansatz
$Q=\kappa\sigma_3$ (an imaginary part of $Q$ can be absorbed by a shift
in frequency and will be inessential for us, while the real part $\kappa$ 
defines the scattering mean free path), where the Pauli matrix reflects
the causal increment $i \delta \sigma_3$. Substituting it into the
equation above and tracing over spin indices we obtain 
\begin{align}
\label{app:scba}
\kappa
&=
-\frac{w}{2} 
{\rm Im}
\int (dk)\, 
\left(
\frac{2}{3}
\frac{\omega+i\kappa}{(\omega+i\kappa)^2-k^2}
+
\frac{1}{3}
\frac{1}{\omega+ i\kappa}
\right).
\end{align}
The detailed solution of the equation then leads to a function $\kappa=\kappa(w,\omega,\Lambda)$. However, for our purposes, it will be sufficient to neglect these parametric dependencies, as we are working at fixed values of the argument parameters.


{\it Soft mode action:---}We next recall that the
 action of the matrix integral, possessed a continuous symmetry under
 $\textrm{Sp}(4R)$ transformations $T_\chi$ block diagonal in helical space.
 Following the standard Goldstone mode paradigm, we promote these
 transformations to modes $T_{\chi,\textbf{x}}$ fluctuating in real space, and
 consider the generalized mean field configurations $\kappa T_\chi\sigma_3^{\rm
 ra}T_\chi^{-1}\equiv 
 Q_\chi$. Here,  the Goldstone modes $Q_\chi$, which
 are the effective degrees of freedom of the nonlinear $\sigma$-model of
 localization theory for symmetry class AI, take values in the coset space
 $\textrm{Sp}(4R)/\textrm{Sp}(2R)\times \textrm{Sp}(2R)$\cite{Wegner1979}. (The
 division of the factor group accounts for the fact that transformations
 commuting with $\sigma_3$ are unbroken symmetries, much like
 rotations around the magnetization-axis in a ferromagnetic material.) The
 symmetry under time reversal discussed above is inherited by the $Q$'s in the
 form $Q_\chi^T=\sigma_2^{\rm tr}Q_\chi\sigma_2^{\rm tr}$. 
 
 With these degrees of freedom,   the effective action  becomes a sum of two
 helical contributions ${\cal Z}=\int D(T_+,T_-) e^{-S[T_+]-S[T_-]}$, with  
  \begin{align}
\label{app_eq:log_action}
S[T_\chi]
&=
-\frac{1}{2}{\rm Tr}\ln\left( \omega - \chi\slashed{k} 
+  i\kappa T_\chi\sigma_3T_\chi^{-1}
\right).
\end{align} 
Where  `${\rm Tr}$' is a trace  over internal degrees of freedom and space.  
A naively applied similarity transformation under the `trace-log',
${\rm Tr}\ln\left( \ldots \right)\to {\rm Tr}\ln\left( T_\chi^{-1} (\ldots) T_\chi \right)$,  
would lead to a putatively equivalent expression   
in terms of 
$\slashed{A}_\chi \equiv S_i A_{\chi,i}  \equiv  S_i T_\chi^{-1}\partial_i T_\chi$. However, this operation is premature because the similarity transformation of the UV singular Dirac like operator under the trace is subject to the parity anomaly~\cite{PhysRevD.29.2366}. One way to handle the situation~\cite{ALTLAND2002283,PhysRevB.93.075113} is to consider 
\begin{align}
\label{app_eq:log_action}
S[T_\chi]
&=
-\frac{1}{2}{\rm Tr}\ln\left( \omega +  i\kappa\sigma_3 -\chi\slashed{k} - i\chi\slashed{A}_\chi
\right)
-
S_\eta
[T_\chi],
\end{align}
where 
 $S_\eta[T_\chi]
=
\frac{1}{2}{\rm Tr}\ln\left( i\eta\sigma_3 -\chi\slashed{k} - i\chi\slashed{A}_\chi
\right)$
with $\eta\to 0$ is introduced to regularize UV divergences at large $\slashed{k}$.

\emph{Gradient expansion:---}The final construction step is the expansion up to
third order in the spatial gradients contained in  $\slashed{A}_{\chi}$.
Inspection of the logarithm shows that the expansion actually is in the
combination $G_\chi\slashed{A}_\chi$, where   
$G_\chi\equiv (\omega +  i\kappa\sigma_3 - \chi\slashed{k})^{-1}$ is
the photon propagator coupled to the SCBA scattering self energy. To proceed, it
is convenient to decompose $G_\chi$ into longitudinal and transverse modes.  
Introducing the projection operators $\pi^{\rm l}_\bold{k}
=
\bold{k}\bold{k}^T/k^2$ and  
$\pi^{\rm t}_\bold{k}=
1-\pi^{\rm l}_\bold{k}$, it is  straightforward to verify that
$G_{\chi k}
=
G_{\textrm{t},\chi,\textbf{k}}\, \pi^{\rm t}_\bold{k}
+
G_{\textrm{l}}\, \pi^{\rm l}_\bold{k}$, with
\begin{align}
  G_{\textrm{t},\chi,\textbf{k}}
  &=
  \frac{
  \omega + i\kappa \sigma_3
  + \chi\slashed{k}
  }{
      (\omega+  i\kappa \sigma_3 )^2-k^2
  },
  \qquad
  G_{\textrm{l}}
  =
  \frac{
  1}{
  \omega + i\kappa \sigma_3 }.
  \end{align}
With this decomposition, it is a matter of a straightforward if tedious
computation to first take the traces over the three-dimensional spin
representation space, and then integrate over momenta. As a result, we obtain a
symmetric second order $(\partial_i \partial_i)$ and a fully antisymmetric third
order $(\epsilon_{ijk}\partial_i \partial_j \partial_k)$ combination of
derivatives acting on the slow fields of the theory, $T$. The first of these is
weighted by  products $G_\textrm{t}G_\textrm{t}$ and
$G_\textrm{t}G_\textrm{l}$, whose momentum integral yields the coefficient
$\sigma$ stated in the main text. At third order we obtain a coefficient
$G_\textrm{t}G_\textrm{t}G_\textrm{l}$, the product of three propagators defining
the `triangle graph' of a Chern-Simons term (cf. Fig.~\ref{fig:FeynmanDiagram}
and Ref.~\cite{PhysRevD.29.2366}). One of these is a longitudinal mode, 
highlighting the importance of the non-dispersive `flat band' in the 
localization physics of the spin-$1$ Weyl system. The momentum integral over 
these propagators yields the universal coefficient multiplying Eq.~\eqref{eq:chern_simons}, 
required on general grounds for a Chern-Simons action. 
For a full exposure of the lengthy calculations leading to these results 
we refer to the Supplementary Note 2.

{\it Helicity mixing:---}We finally include  the helicity-mixing $\tau_1\slashed{a}_Z$ 
operator into the analysis. With $\slashed{A}_Z\equiv T^{-1}\tau_1 \slashed{a}_ZT$ and 
block diagonal $T=T_+ \oplus T_-$, we consider the unitarily transformed 
from a locally varying impedance, 
\begin{align*}
S[T]
&=
-
\frac{1}{2}{\rm Tr}\ln\left( \omega +  i\kappa\sigma_3 -\tau_3\slashed{k} + \slashed{A}_Z
+\dots\right),
\end{align*}
where the ellipses denote $T$-dependent contributions previously accounted for in the gradient expansion. We expand to second order in $\slashed{A}_Z$, and average  over the distribution given in Table~\ref{table:1}, 
to arrive at the effective quadratic coupling action 
\begin{align*}
&S_\textrm{h}[T]=
-\frac{\gamma}{4}\int d^3x\,
{\rm tr}\left(
TG_{\bold{x},\bold{x}}T^{-1}
\tau_1 S_i
TG_{\bold{x},\bold{x}}T^{-1}
\tau_1 S_i
\right)
\nonumber\\
&=
\left(\frac{2\kappa}{w}\right)^2 \times \frac{3\gamma}{4}\int d^3x\,
{\rm tr}\left(
Q
\tau_1 
Q
\tau_1 
\right),
\end{align*}
where $G=G_+\oplus G_-$. In passing from the first to the second line, we used that the Green functions $G_\chi(\textbf{x},\textbf{x})=\int (dk) G_{\chi,k}$ solve the mean field equation, i.e. $G(\textbf{x},\textbf{x})=(- 2\pi i\kappa/w) \sigma_3$.
Tracing over helical space we then arrive at Eq.~(8) in the main text, with the coupling constant $\Gamma_\textrm{h}$ specified in Table~\ref{table:1}.

{\it Limits of applicabilty:---}We conclude with some  comments on
  the validity of the effective field theory. The discussion of the
  applicability of the nonlinear sigma model in the strong scattering regime has
  a long history and remains a topic of debate. (For a recent critical
  contribution, see e.g. Ref.~\cite{Zirnbauer2023}). However, there appears to
  be a consensus that in the absence of topological effects  
  the Anderson transition follows a one-parameter scaling hypothesis adequately
  described by a renormalization group approach to the sigma model in $2 +
   \epsilon$ dimensions. 
  
Topological terms in these theories stand on a more solid footing as their
presence follows entirely on the basis of  symmetries~\cite{Ryu:2010fk},
independent of specific details such as disorder strength. However, by the same
principle, effects characterizing ``symmetry protected topological phases''
generally depend on the presence
of certain symmetries. For example, magnetic fields compromise
topological protection in  time-reversal invariant topological insulators,
technically by gapping
out topological terms. An analogous scenario is realized here where topological
protection is guaranteed, unless helicity symmetry is broken in a manner similar
to spin rotation invariance breaking in quantum theories.

{\bf Acknowledgement.} We thank Antonio Zelaquett Khoury and Felipe Pinheiro 
for stimulating discussions and helpful comments on the manuscript.  
Financial support by Brazilian agencies CNPq and FAPERJ and the Deutsche Forschungsgemeinschaft (DFG) project grant 277101999
within the CRC network TR 183 (subproject A03) is acknowledged.

{\bf Data availability.} Data sharing not applicable to this article as no datasets were generated 
or analysed during the current study.

{\bf Author contributions.} T.~M. and A.~A. contributed to the development of ideas, calculations, discussion of results
and writing of the manuscript.

{\bf Competing interests.} The authors declare no competing interests.


\end{document}